\documentclass{article}
\usepackage{spconf,amsmath,graphicx}

\usepackage{caption}
\usepackage{subcaption}
\usepackage{enumitem}
\usepackage{inconsolata}

\title{On the pragmatism of using binary classifiers over data intensive neural network classifiers for detection of COVID-19 from voice}
\name{Ankit Shah$^{\star}$, Hira Dhamyal$^{\star}$, Yang Gao$^{\star}$, Daniel Arancibia$^{\dagger}$,  Mario Arancibia$^{\dagger}$,  Bhiksha Raj$^{\star}$, Rita Singh $^{\star}$ } 
\address{$^{\star}$Carnegie Mellon University, Pittsburgh PA USA \\ $^{\dagger}$ Merlin Inc, Santiago, Chile}

\begin{document}
\ninept
\maketitle
\begin{abstract}

Lately, there has been a global effort by multiple research groups to detect COVID-19 from voice. Different researchers use different kinds of information from the voice signal to achieve this. Various types of phonated sounds and the sound of cough and breath have all been used with varying degree of success in automated voice based COVID-19 detection apps. In this paper, we show that detecting COVID-19 from voice does not require custom made non-standard features or complicated neural network classifiers rather it  can be successfully done with just standard features and simple binary classifiers. In fact, we show that the latter are not only more accurate and interpretable but also more computationally efficient in that they can be run locally on small devices. We demonstrate this on a human-curated dataset of over 1000 subjects, collected and calibrated in clinical settings.
\end{abstract}
\begin{keywords}
 Detecting Covid-19 from Voice, Binary classifier, CNN for voice analysis. 
\end{keywords}
\section{Introduction}
\label{sec:intro}

The COVID-19 pandemic no longer needs an introduction, having sickened -- and killed -- millions of people worldwide in just a few months \cite{worldometer}. One of the primary weapons on our arsenal to combat this pandemic is isolation -- identifying infected or potentially infected individuals, and isolating them, to prevent them from infecting others. Identifying such individuals requires tests, and testing capacity for Covid-19 remains inadequate, with often unacceptable wait times during which an individual may spread the infection \cite{cvdtesting}. Consequently the search remains ongoing for rapid tests that can be quickly dispensed with minimal logistical challenges. Voice-based testing has been identified as one such possibility \cite{cvdvoicedetector}.

COVID-19 primarily manifests as an illness of the respiratory tract \cite{asiaee2020voice}, although other organs too are affected. In symptomatic patients, it affects the lower and upper respiratory tract \cite{9414201_soham}. While, by the very widespread nature of the pandemic and the shortage of testing capacity, the actual number of infected people and fraction of this population that exhibited specific symptoms can never be known precisely, various reports suggest that between 60\% and 80\% of symptomatic patients develop a dry cough \cite{song2021confronting}, and roughly a third develop a wet cough \cite{song2021confronting}. A common symptom is respiratory distress , and fluid accumulation in the lungs is frequently reported even in early-stage patients \cite{suess2020gross}. 

Naturally, it is to be expected that these impairments will manifest in the patient's vocalizations -- both in the form of coughing and in changes to their voice.  COVID-19 affects the three stages in voice production: respiration, phonation and the underlying resonant system \cite{asiaee2020voice}. The reduced lung capacity affects exhalation, affecting the mechanisms for the sub-glottal pressure and disrupting normal phonation \cite{asiaee2020voice}. In a self-assesment study, COVID patients reported difficulty producing certain voiced sounds and noticed changes in their voice \cite{lechien2020features}. 

Consequently, a number of research groups around the world have initiated efforts on attempting to diagnose potential Covid infections from recordings of vocalizations \cite{9414530_soham,9414201_soham}. While most groups have focused on cough sounds \cite{brown2020exploring, imran2020ai4covid, bagad2020cough} as they are a frequent symptom of Covid-19, several groups have also considered other vocalizations, such as breathing sounds \cite{brown2020exploring,10.1145/3394486.3412865} extended vowels \cite{RitaSinghTalkCambridge:2020, pinkas2020sars,schuller2021covid}, and counts. Yet other teams have analyzed free-form speech such as those obtainable from YouTube recordings\cite{shukla2021covid}. The general consensus appears to be that cough sounds \cite{10.1145/3394486.3412865}, and possibly speech sounds do provide discriminative information about the presence of COVID-19 . 

Promising results have also been obtained by using estimates of vocal fold oscillations \cite{9414530_soham,9414201_soham}. In all of these efforts, researchers have ubiquitously used neural networks for classification. Neural networks however require larger amounts of data for robust training. In the case of COVID-19 many groups have mentioned the difficulty in getting properly curated data in sufficient amounts that are needed for neural network training. Data insufficiency has cast a doubt on the basic premise of whether voice samples can accurately be used to detect COVID-19 from voice.  The question that this paper explores is : if we use classifiers for which data sufficiency is not such a critical issue, would the detection of COVID-19 from voice be positively affected. If so, \textit{i.e.} if simpler and less data intensive classifiers were able to achieve good results for detecting COVID-19 from voice then we could exploit this to create much faster, more robust and more accurate and more portable COVID-19 detection applications. 

While the premise of this paper is simple, this exploration is nevertheless crucial given the difficulty in obtaining clinical data not only for COVID-19 but also for many other health conditions. In this paper, we explore several simple classifiers and compare and contrast them to neural network based classifiers using standard features available publicly for voice analysis/speech applications. The latter is a choice that has been deliberately made in order to avoid the extra effort of inventing new features in a time-critical pandemic situation. It brings out the value of confidently relying on time-tested features and classifiers for detecting novel entities from voice. It brings out the importance of doing this exercise first before presuming that the situation is complex and that it requires complex solutions. Sometimes complex challenges do have easy to implement and simple solutions. It is purpose of this paper to highlight this.

The rest of the paper is organized as follows. In Section 2, we discuss the work that has been done for detecting COVID-19 from voice from multiple perspectives. In Section 3, we describe the features and classifiers that have been used in multiple settings and explored by us in this paper for bringing out our main points. Section 4 describes the our data and experiments. In the section 5, we present our conclusions.

\section{Related Work}
It is by now well known that speech can carry the signatures of numerous ailments \cite{singh2021overview}. Specifically, any ailment that affects any portion of the vocal apparatus may be expected to leave its signatures in voice.  Covid-19, which affects the lungs and muscles among other things, falls in this category. Consequently, many researchers have developed voice based analyzers for the detection of Covid-19 \cite{elizalde2021covid,9414201_soham,bagad2020cough,schuller2021covid,imran2020ai4covid,despotovic2021detection}. So far, the research has focused on two aspects -- the collection and curation of reliable data in possibly non-clinical settings (e.g. crowd-sourced settings) and  the determination of which vocal sounds might be most discriminative for the detection of Covid-19. Examples include Chloe et.al \cite{brown2020exploring} who collect cough and breathing samples from people who self-report as being infected with Covid. Although their data set is relatively large, the correlation between the true Covid-19 status of the speaker and their reported status is weak, compared to fully clinically validated data. They run SVM and Logistic Regression classifiers with handcrafted spectral features, but the results are unreliable due to the  errors that can be inadvertently or deliberately introduced in the process of self-reporting, as mentioned above.

The study by \cite{pinkas2020sars} analyzes selected vowel sounds from speech recordings of counting (e.g. from 50 to 80), collected over celluar phones, using neural networks. It concludes that the sounds of counting, in particular, the /z/ sound, are useful for distinguishing Covid. \cite{asiaee2020voice} report results on data collected in clinical settings and discuss how various acoustic parameters of vowel sounds like /aa/ change between the infected and non-infected people. The parameters analyzed include fundamental frequency, jitter and shimmer, HNR etc.

The AI4Covid Application (described in \cite{imran2020ai4covid}) performs analysis on 70 samples from subjects who were positive for Covid-19 and 96 Covid-negative samples and reports an 89\% F-1 score for the detection of Covid-19. However, since the data size used is small, the authors do not perform cross-validation experiments.

The study in \cite{pinkas2020sars} performs analysis of 29 Covid-positive and 59 Covid-negative patients' data with the sounds /ah/, /z/ and  counting. Their model extracts a feature representation using a transformer RNN network architecture with an added  speaker dissimilarity loss. The final score for their model is obtained using a libSVM-based classifier, yielding an AUC of  0.78, obtained with a leave-one-out based cross-validation strategy, owing to the small dataset size.  \cite{bagad2020cough} performs analysis on a relatively large-scale dataset comprising 376 positive cough samples, and 663 negative cough samples. The analysis uses a ResNet-based CNN architecture for making a binary prediction (Covid vs non-COVID) and yields an AUC of 0.68 on their dataset. However, their method uses a relatively high portion (90\%) of the total data for training.

The Coswara dataset \cite{sharma2020coswara} comprises breathing, cough, vocalized sounds, which include the vowels /e/, /i/, /a/, and shallow and fast deep breathing sounds collected from  Covid positive and negative patients. They attempt to classify 9 sound categories, with features such as spectral contrast, MFCC, spectral roll-off, zero-crossing rate, spectral bandwidth, spectral flatness etc.,  and achieve an accuracy score of 66.74\% on their test set. The low accuracy presumably stems from the split of the (small amount of) data.

\cite{10.1145/3394486.3412865} collected cough and breath sounds from 7000 unique users, but with only 235 covid-positive samples. Using a combination of features, including standard ones such as duration, tempo, RMS energy, spectral centroids and Vggish based features,  the approach yields an AUC of 0.8 for COVID vs non-COVID classification.

In all cases, the  reported performance is often either poor or unreliable, due to the unavailability of a sufficiently large and reliable dataset. The research community has yet to work on a dataset of sufficient size, with a high number of both covid-positive and covid-negative speakers, would be sufficiently large to perform reliable k-fold cross-validation with a large-parameter neural network classifier. Our goal in this paper is to find the cusp where the need for large amounts of data is minimized while the classification accuracy is not compromised. For this, we explore standard less-data intensive classifiers to find the best balance of dataset size, features and classifiers, to achieve our goal.



\section{Features and Classifiers}
\label{sec:data}

The primary premise of our paper is that given only limited scales of cleanly labeled data, performance comparable with what is currently state-of-art is achievable with simpler setups using canned feature extractors that are not specifically optimized for the task, and simple classifiers. For this to hold, it is necessary to select the appropriate set of features and classifiers. We discuss our choices below.

\subsection{Features Extraction} \label{ssec:feats}
We extract a multitude of features from voice using tools that are not specifically optimized on our test data, or for the Covid task. 

We use the OpenSMILE toolkit \cite{eyben2010opensmile}, with the inbuilt \emph{emobase20-10.conf} configuration file. This extracts features such as intensity and loudness, Mel-frequency cepstral coefficients (MFCC), Linear Predictive Coding (LPC) features, pitch,  and voice quality features such as jitter and shimmer. In total, a 1582-dimensional feature vector composed of of such features is extracted  per audio recording. We call this the ``Ovec feature'' in Section \ref{sec:results}. We also use features extracted by the pre-trained ``PANN'' model \cite{kong2019panns}, which has been trained on audio datasets,  and from the ``VGG''  model that has been pre-trained model on the VGG dataset \cite{45611}. We also extract feature representations from the PASE model \cite{pase, pase+}. In all cases, we use the default settings for these tools to extract features. 
In all cases, the frame (or chunk) level features derived for each recording are averaged to obtain an utterance level feature.

We also extract frame-level spectral features for the audio using Librosa \cite{mcfee2015librosa}, using an analysis frame width of 2048 samples, with a frame shift of 512 samples (once again these are default values; other variants were not seen to provide any benefits). The spectral features include the zero-crossing rate,  spectral centroids, spectral roll off frequency, tempo, root-mean-square energy, and 20-dimensional MFCCs as well as their differences (velocity) and double-differences (acceleration).  These too are averaged across the recording to obtain a single vector for each recording.
For all of these spectral features except tempo, we also extracted several statistical features such as min, max, mean, RMS value,  median, inter-quartile range, first, second and third quartile limits, standard deviation, skew, and kurtosis to capture their distribution in the recording. This collectively returned a 14-component statistical feature per spectral feature. We also compute these 14 statistical features for each of the 20 MFCC coefficients, thus obtaining a 280 dimensional feature vector. In total, we finally obtain 833 spectral features, which we refer to as ``custom'' features in section \ref{sec:results}.  In addition, we also extracted features using YAMNet\cite{yamnet}  as well as the Open-L3 \cite{Cramer:LearnMore:ICASSP:19} based feature extractor for each voice recording to compare their performance with Vggish \cite{hershey2017cnn} based feature extractor. 


\subsection{Classifier Description}
To verify our original premise, we must compare simple binary classification using traditional techniques with neural network classifiers. We outline our choices below.  \vspace{-2mm}
\subsubsection{Neural models}   \label{ssec:neural}
To ensure fair comparison, we have used neural architectures that have previously been reported to be among the most successful on the Covid task.  The neural architectures we tested combined a variety of state-of-art neural feature extractors, including PANN \cite{kong2019panns}, Vggish \cite{45611}, Open-L3 \cite{Cramer:LearnMore:ICASSP:19}  and the PASE model \cite{pase,pase+}, with additional layers of LSTMs, GRUs and CNNs.  We considered both, freezing the pre-trained feature extraction and only tuning the final LSTM, GRU or CNN, vs. updating both -- fine-tuning the feature extractor and learning the upper-layer classifiers. We also trained end-to-end systems that directly took in spectrographic features derived from the speech signal to perform the classification.  Since the actual classification is binary, our final classification layer, in all cases, was a simple binary softmax layer. \vspace{-2mm}

\subsubsection{Simple binary classifiers} \label{Ssec:binary}
We also tested a number of ``simple'' (non-neural network based) binary classifiers including RandomForests (RF), Support Vector Machines (SVM) with various kernels, and Logistic Regression (LR) in order to perform classification. \vspace{-1mm}

\section{Data and Experiments} \vspace{-2mm} 
\label{sec:experiments}

\subsection{Dataset and its description}
A key challenge in many of the previous studies was the data -- data that were trustworthy and medically reliable were generally small, while those that were collected at large scale were generally self-reported and unreliable. To address this issue, we require a dataset that is both reasonably large, and reliably annotated.
We used a dataset collected under clinical supervision and curated by Merlin Inc., a private firm in Chile. 

The data consists of recordings from 421 covid-positive and 989 covid-negative unique subjects. The Covid-positive or negative labels were obtained through laboratory tests. Subjects were asked to record themselves saying the elongated vowels /AH/, /UW/ and /IY/, and speaking the Spanish alphabet (a-z), counting from 1-20 (in Spanish) and producing coughs. 
Subjects were recorded in a quiet room environment over a smartphone, and sampled at 8khz.
The total data comprised 17.5 hours of covid-positive and 20.5 hours of covid-negative recordings, for a total of over 37 hours of data.

\begin{figure}[!t]
\centering
\includegraphics[width=8.5cm, height=3.9cm]{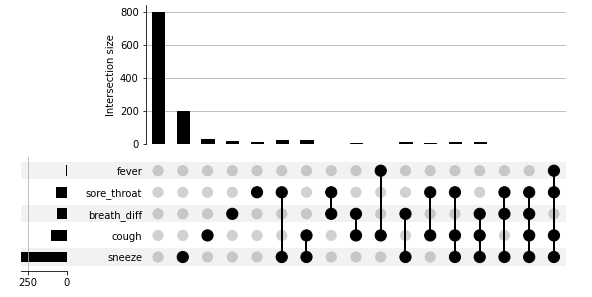}
\includegraphics[width=8.5cm, height=3.9cm]{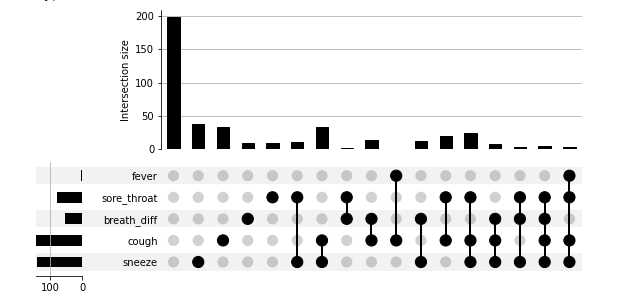}
\caption{The figure shows the distribution of symptoms in Covid-19 positive patients (top) and negative individuals (bottom) (breath-diff stands for breathing difficulty). The rows represents each symptom whereas the column represents the frequency of occurrence of a symptom which are marked by black dots.}\label{fig:symptoms}
\end{figure}

A number of covariates that could potentially influence the classification of Covid were also recorded along with the data.  Subjects usually have symptoms of coughing, sneezing, breathing difficulties etc. These were recorded to account for symptomatic and asymtomatic Covid-19 diagnosis. Figure \ref{fig:symptoms} shows the prevalence of different symptoms for covid-positive and covid-negative subjects. 

The occurrence of Covid is also reportedly related to the age of the subjects; these too were recorded. Additional metadata about preexisting conditions such as smoking preference, asthma and detailed comments about the speakers health at the time were also recorded. Table \ref{tab:age_table} gives the distribution of subjects by age. For the purpose of the study we have  broadly categorized the ages into four groups as shown. Table \ref{tab:asthma_table} shows the prevalence of Asthma and smoking among the subjects.

\begin{table}[!t]
    \centering
    \begin{tabular}{|c|c|c|c|c|}
        \hline 
          age $\leq$ 30 & 30 $<$ age $\leq$ 40 & age $>$ 40 & N/A \\
        \hline \hline
             192 & 209 & 225 & 186\\
            \hline
    \end{tabular}
    \caption{Distribution of subjects by age in years. N/A represents subjects for who age was not reported. Only subjects for whom metadata were available are listed.}
    \label{tab:age_table}
        \vspace{-4mm}
\end{table}

\begin{table}[!t]
    \centering
    \begin{tabular}{|c|c|}
    \hline
        \textbf{Has Asthma} & \textbf{Is Smoker} \\
        \hline \hline
             50 & 234\\
             \hline
    \end{tabular}
    \caption{Asthma and smoker statistics}
        \vspace{-4mm}
    \label{tab:asthma_table}
\end{table}
\vspace{-2mm}
\subsection{Experiments}
\label{sec:results}
Our objective is to achieve competitive Covid-19 classification performance using simple models on audio recordings. All classification experiments below were run only the voice samples, except where specified otherwise.

For our comparative experiments we ran a variety of both nominally high-performance neural-architectures described in Section \ref{ssec:neural} and ``simple''  classifiers as outlined in Section \ref{Ssec:binary}. For the latter, in addition to the signal features (``Ovec'' and ``Custom'') described in Section \ref{ssec:feats}, we also used pre-trained neural feature extractors, as these were fixed for these models and are not trained or optimized on the data.

All classifiers were evaluated using 5-fold cross-validation, ensuring that within any fold all of a subject's recordings appeared either in the training fold or the test fold, but not both (thereby ensuring that the model does not inadvertently just learn speaker characteristics). We run grid search on all these classifiers to find the best hyper-parameters.

The models were trained using all audio samples. To classify a subject, all six of their voice recordings were classified, and their classifier scores averaged. The final classification performance was quantified by the area under the DET curve that charts true positive rate (precision) vs false positive rate. The ideal AUC is 1.0.

Among the neural architectures, the best performance was consistently obtained with unadapted, pre-trained PASE features. Among models using PASE features, LSTMs and GRUs achieved AUCs of 0.62 and 0.65 respectively, where as CNNs achieved a much higher AUC (see Table \ref{tab:my_label}).  Among the ``simple'' models Random Forests consistently outperformed other classifiers under nearly all settings, and resulted in the best overall performance by a significant margin, compared to the other simple models.

Table \ref{tab:my_label} shows a more detailed set of results.  As we can see the best overlal performance is obtained by Random Forests working on a combination of Ovecs, custom features and features derived by YAMNet, clearly outperforming the best optimized neural model. RFs achieve an AUC of 0.86 (which corresponds to a precision of 0.9 at a recall of 0.5), compared to the best AUC of 0.77 achieved by any optimized neural-network model.

\begin{table}[h]
    \centering
    \resizebox{1.0\columnwidth}{!}{
    \begin{tabular}{|c|c|c|c|}
    \hline 
       \textbf{Classifier}  &  \textbf{Feature}  & \textbf{AUC} \\ 
       \hline \hline
        Random Forest & Ovec & 0.76 \\
        Random Forest & PANN & 0.78    \\
        Random Forest & vggish & 0.59   \\
        Random Forest & Open-L3 & 0.64   \\
        Random Forest & YAMNet & 0.72   \\
        Random Forest & Custom  Feat & 0.82  \\ 
        Random Forest & Ovec + Custom  Feat & 0.84  \\
        Random Forest & Ovec + Custom Feat + vggish & 0.82  \\
        Random Forest & Ovec + custom feature + YAMNet & 0.86  \\ 
        CNN & Spectrogram & 0.69   \\
        CNN & PASE & 0.77     \\ 
        \hline
    \end{tabular}
    }
    \caption{Comparison of the performance of different classifier-feature combinations.}
    \vspace{-2mm}
    \label{tab:my_label}
\end{table}

Investigating further, Table \ref{tab:by_sound} presents the classification performance obtained using each of the six individual sound types. Once again, for every sound  type the simple random-forest classifier outperformed the more complex CNN. Interestingly, the best classification is obtained using vowel sounds, rather than coughs.

\begin{table}[]
\begin{tabular}{|l|c|c|c|c|c|c|}
\hline
Audio & /AH/ & /IY/ & /UW/ & a-z & 1-20 & cough \\ \hline
CNN (PASE) & 0.68 & 0.68 & 0.66 & 0.7 & 0.68 & 0.65 \\ \hline
RF & 0.79 & 0.82 & 0.80 & 0.78 & 0.76 & 0.78 \\ \hline
\end{tabular}
\caption{AUC  on individual sound types using PASE features.}
\label{tab:by_sound}
\end{table}

Tables \ref{tab:my_label} and \ref{tab:by_sound} prove our hypothesis that a binary classifier such as Random Forest can perform better for Covid-19 detection than CNN (or other neural) counterparts under limited-data settings. Pre-trained neural feature extractors do continue to provide benefits, although  standard signal-based features are also equally effective.



We now consider whether additional covariates may be used to condition and improve the performance. Although the performance of the classifier differed by age group (performing best on the $>40$ age group), explicit introduction of age as a feature did not affect the performance of either classifier significantly, presumably because this information is already present in the voice.

It is generally assumed that the symptoms of covid may be useful to identify the illness. Figure \ref{fig:symptoms} shows that some symptoms, e.g. coughs, are indeed more prevalent among Covid patients. However, as also evident from Figure \ref{fig:symptoms}, subjects are predominantly asymptomatic and this dominates analysis. For reference, classification based only on symptoms (using a Random Forest classifier) achieves only a precision of 0.4 at a false positive rate of 0.1 and a recall of 0.52, and a precision of only 0.1 at a false positive rate of 0.01. 

Inclusion of symptoms into the voice-based classifier has negligible effect on the overall classifier performance, and does not change the relative ordering of the classifiers -- the RandomForest classifier still outscores the CNN.   Figure \ref{fig:symptoms} shows that most of the symptoms (which are commonly associated with Covid)  are not discriminative in our dataset. Only cough is somewhat discriminative; nonetheless it does not add substantially to the evidence in the audio recordings (which include cough). Overall, the change in performance from inclusion of symptoms is not statistically significant. 

Having a  history of asthma or smoking might put people at a higher risk of COVID \cite{asthma}. Our experimental results indicate asthma is more correlated to COVID-19 than smoking: the recall of the RF classifier is 66\% for smokers, vs 80\% for asthmatics. However, providing this information explicitly did not modify overall performance for either classifier, presumably because of the limited number of subjects for which the information was available.

An added benefit of using simple classifiers like Random Forests is the ability of interpreting the results. 
Analyzing the 40 most important features chosen by the RF model, we find that 12 of these relate to voicing, and 13 relate to line-spectral frequencies, which capture the resonances of the vocal tract.  This finding is in line with the previous work on COVID-19's effect on voice which shows how the virus can cause  sore throat, damage vocal folds and end up effecting the voice and its resonances. The remaining features largely capture various characterizations of the spectrum, as expected, indicating that Covid influences the spectral structure of the signal.

\section{Conclusion}\vspace{-1mm}
\label{sec:discuss}
Motivated by the urgent need to have alternative methods to augment the medical tests, we designed the experimental setup for Covid-19 analysis using a wide variety of voice samples. One of the key challenges of such urgent situations is that there is generally little data to develop high-quality classifiers.  We demonstrate that in such settings even simple conventional binary classifiers working with appropriately selected ``canned'' features can significantly outperform data-intensive neural models that generally achieve much higher performance given sufficient data, enabling one to quickly ramp up automated diagnosis technologies.

Specifically for Covid-19, our analysis also provides some additional insights. It demonstrates that vowel sounds may, in fact be as effective as cough sounds to detect Covid, and that voice-based diagnosis, when properly designed, may actually provide better preliminary screening that tests based on symptoms alone, particularly when most subjects are asymptomatic. Our choice of models also provides the ability of interpreting the results. Specifically using Random Forests, we find that the key features that distinguish covid relate to voicing and the resonances of the vocal tract.

\section{Acknowledgements}\vspace{-1mm}
This material is based upon work supported by the Defence Science and Technology Agency,  Singapore under contract number A025959. Its content does not reflect the position or policy of DSTA and no official endorsement should be inferred.




\bibliographystyle{IEEEbib}
\bibliography{strings,refs}

\end{document}